\documentclass{aa}  

\usepackage{graphicx}
\usepackage{txfonts}
\usepackage{xcolor}
\usepackage{hyperref}
\hypersetup{colorlinks, linkcolor=blue, citecolor=blue, urlcolor=blue}
\usepackage{amsmath}
\usepackage{ulem}
\usepackage{soul}
\usepackage{enumitem}

\newcommand{\pivec}{\mbox{\boldmath $\pi$}}
\newcommand{\muvec}{\mbox{\boldmath $\mu$}}
\newcommand{\xivec}{\mbox{\boldmath $\xi$}}

\newcommand{\te}{t_{\rm E}}
\newcommand{\thetae}{\theta_{\rm E}}

\newcommand{\pie}{\pi_{\rm E}}
\newcommand{\pien}{\pi_{{\rm E},N}}
\newcommand{\piee}{\pi_{{\rm E},E}}
\newcommand{\dl}{D_{\rm L}}
\newcommand{\ds}{D_{\rm S}}

\newcommand{\xien}{\xi_{{\rm E},N}}
\newcommand{\xiee}{\xi_{{\rm E},E}}

\definecolor{brown}{rgb}{0.59, 0.29, 0.0}
\definecolor{darkgreen}{rgb}{0.0, 0.42, 0.24}
\definecolor{darkblue}{rgb}{0.01, 0.31, 0.59}
\definecolor{blue}{rgb}{0.0,0.0,1.0}
\definecolor{green}{rgb}{0.0,1.0,0.0}

\def\eqalign#1{\null\,\vcenter{\openup\jot
        \ialign{\strut\hfil$\displaystyle{##}$&$
        \displaystyle{{}##}$\hfil \crcr#1\crcr}}\,}

\begin{document} 

\title{
KMT-2021-BLG-0240: Microlensing event with a deformed  planetary signal}

\author{
     Cheongho Han\inst{\ref{01}} 
\and Doeon~Kim\inst{\ref{01}} 
\and Hongjing~Yang\inst{\ref{02}}
\and Andrew Gould\inst{\ref{03},\ref{04}} 
\and Youn~Kil~Jung\inst{\ref{05}} 
\and Michael~D.~Albrow\inst{\ref{06}} 
\and Sun-Ju~Chung\inst{\ref{05}} 
\and Kyu-Ha~Hwang\inst{\ref{05}} 
\and Chung-Uk~Lee\inst{\ref{05}} 
\and Yoon-Hyun~Ryu\inst{\ref{05}} 
\and In-Gu~Shin\inst{\ref{05}} 
\and Yossi~Shvartzvald\inst{\ref{07}} 
\and Jennifer~C.~Yee\inst{\ref{08}} 
\and Weicheng~Zang\inst{\ref{02}} 
\and Sang-Mok~Cha\inst{\ref{05},\ref{09}} 
\and Dong-Jin~Kim\inst{\ref{05}} 
\and Seung-Lee~Kim\inst{\ref{05}} 
\and Dong-Joo~Lee\inst{\ref{05}}
\and Yongseok~Lee\inst{\ref{05}} 
\and Byeong-Gon~Park\inst{\ref{05}} 
\and Richard~W.~Pogge\inst{\ref{04}}
\\
(The KMTNet Collaboration) \\
}

\institute{
       Department of Physics, Chungbuk National University, Cheongju 28644, Republic of Korea  \\ \email{cheongho@astroph.chungbuk.ac.kr}              \label{01} 
\and   Department of Astronomy, Tsinghua University, Beijing 100084, China                                                                             \label{02} 
\and   Max Planck Institute for Astronomy, K\"onigstuhl 17, D-69117 Heidelberg, Germany                                                                \label{03} 
\and   Department of Astronomy, The Ohio State University, 140 W.  18th Ave., Columbus, OH 43210, USA                                                  \label{04} 
\and   Korea Astronomy and Space Science Institute, Daejon 34055,Republic of Korea                                                                     \label{05} 
\and   University of Canterbury, Department of Physics and Astronomy, Private Bag 4800, Christchurch 8020, New Zealand                                 \label{06} 
\and   Department of Particle Physics and Astrophysics, Weizmann Institute of Science, Rehovot 76100, Israel                                           \label{07} 
\and   Center for Astrophysics~|~Harvard \& Smithsonian, 60 Garden St., Cambridge, MA 02138, USA                                                       \label{08}
\and   School of Space Research, Kyung Hee University, Yongin, Kyeonggi 17104, Republic of Korea                                                       \label{09}  
}
\date{Received ; accepted}

\abstract
{}
{
The light curve of the microlensing event KMT-2021-BLG-0240 exhibits a short-lasting anomaly 
with complex features near the peak at the 0.1~mag level from a single-lens single-source model.  
We conducted modeling of the lensing light curve under various interpretations to reveal the nature 
of the anomaly.
}
{
It is found that the anomaly cannot be explained with the usual model based on a binary-lens (2L1S) 
or a binary-source (1L2S) interpretation.  However, a 2L1S model with a planet companion can describe 
part of the anomaly, suggesting that the anomaly may be deformed by a tertiary lens component or 
a close companion to the source.  From the additional modeling, we find that all the features of 
the anomaly can be explained with either a triple-lens (3L1S) model or a binary-lens binary-source 
(2L2S) model  
obtained under the 3L1S interpretation.  However, it is difficult to validate the 2L2S model
because the light curve does not exhibit signatures induced by the source orbital motion and 
the ellipsoidal variations expected by the close separation between the source stars according to 
the model.  We, therefore, conclude that the two interpretations cannot be distinguished with 
the available data, and either can be correct.
}
{
According to the 3L1S solution, the lens is a planetary system with two sub-Jovian-mass planets
in which the planets have masses of 0.32--0.47~$M_{\rm J}$ and 0.44--0.93~$M_{\rm J}$, and they 
orbit an M dwarf host.  According to the 2L2S solution, on the other hand, the lens is a single 
planet system with a mass of $\sim 0.21~M_{\rm J}$ orbiting a late K-dwarf host, and the source 
is a binary composed of a primary of a subgiant or a turnoff star and a secondary of a late G dwarf.  
The distance to the planetary system varies depending on the solution: $\sim 7.0$~kpc according to the 3L1S 
solution and $\sim 6.6$~kpc according to the 2L2S solution.  
}
{}


\maketitle

\section{Introduction}\label{sec:one}

A planetary signal in a microlensing light curve is produced by the source star's approach close 
to or passage through the caustic induced by a planet \citep{Mao1991, Gould+Loeb1992}. In general,
 a planet induces two sets of caustics, in which one set is located near the host of the planet 
(central caustic) and the other set lies away from the host (planetary caustic). The central 
caustic provides an important channel of planet detections for two major reasons. First, the 
planetary signal induced by the central caustic (central anomaly) always appears near the peak of the 
light curve of a high-magnification event \citep{Griest1998}, and thus the time of the signal can 
be predicted in advance, making it possible to densely cover the signal from follow-up observations
\citep{Udalski2005}.  Second, because the central anomaly occurs when the source is greatly 
magnified, the signal-to-noise ratio of the central anomaly is greater than that of the anomaly 
produced by a planetary caustic.  As a result, a significant fraction of microlensing planets 
have been detected via the central anomaly channel, despite the fact that the central caustic is 
substantially smaller than the planetary caustic \citep{Han2006}.

Characterizing a planetary system from an observed central anomaly can often be  fraught with
difficulties in accurately interpreting the anomaly. For the anomaly produced by a planetary
caustic, the planet-host separation $s$ (normalized to the angular Einstein radius $\thetae$) 
can be heuristically estimated from the location of the anomaly in the lensing light curve.  
In contrast, this estimation is difficult for the central anomaly because it appears near the 
peak regardless of the planetary separation.  In addition, the interpretation of the anomaly
is usually subject to the degeneracy between the two solutions with $s$ and $s^{-1}$ 
arising from the intrinsic similarity between the two central caustics induced by planets 
with separations $s$ and $s^{-1}$: close-wide degeneracy. Furthermore, central anomalies can 
be produced not only by a planet but also by a binary companion to the lens \citep{Han2005}, 
and thus distinguishing the two interpretations is occasionally difficult for weak or poorly 
covered signals.

Another difficulty in the interpretation of a central-caustic planetary signal is caused by the 
deformation of the anomaly. The deformation of the central anomaly arises due to various causes. 
The first cause is the multiplicity of planets.  If there exist multiple planets around the Einstein 
ring of the planet host, the individual planets induce their own caustics in the central magnification 
region, causing deformation of the anomaly pattern \citep{Gaudi1998, Han2005}.  Such deformations 
were observed in five lensing events of OGLE-2006-BLG-109 \citep{Gaudi2008, Bennett2010}, 
OGLE-2012-BLG-0026 \citep{Han2013, Beaulieu2016}, OGLE-2018-BLG-1011 \citep{Han2019}, 
OGLE-2019-BLG-0468 \citep{Han2022c}, and KMT-2021-BLG-1077 \citep{Han2022a}.  The second cause 
is the existence of a binary companion to the planet host, that is, planets in binary systems.  
For a planet in a binary stellar system,   orbiting either around one of the two stars of a wide 
stellar binary system (P-type orbit) or around the barycenter of a close binary system (S-type 
orbit), the topology of the critical curve and caustic would be affected by the stellar binarity, 
causing deformation of a planetary signal.  See \citet{Danek2015} and \citet{Danek2019} for the 
detailed variation of the critical curve and caustic in triple-lens systems.  There are four 
events with such deformations, including OGLE-2007-BLG-349 \citep{Bennett2016}, OGLE-2016-BLG-0613 
\citep{Han2017}, OGLE-2018-BLG-1700 \citep{Han2020-1700}, and KMT-2020-BLG-0414 \citep{Zang2021a}.  
Finally, the central-caustic signal can also be deformed by the close companion to the source, as 
illustrated by three lensing events of MOA-2010-BLG-117 \citep{Bennett2018}, KMT-2018-BLG-1743 
\citep{Han2021a}, KMT-2021-BLG-1898 \citep{Han2022b}.

In this paper, we present the analysis of the short-term anomaly that appeared near the peak of 
the high-magnification lensing event KMT-2021-BLG-0240.  The anomaly, which lasted for about 2~days 
at the 0.1~mag level, could not be explained by a usual binary-lens or a binary-source model, and 
we investigate various causes for the deformation of the signal.

The analysis is presented according to the following organization of the paper.  In Sect.~\ref{sec:two},
we give an explanation of the observations conducted to obtain the photometric data analyzed in this 
work and describe the procedure of data reduction. In Sect.~\ref{sec:three}, the details of the anomaly 
feature in the lensing light curve is depicted, and the procedure of the analysis carried out to 
interpret the anomaly is described in detail.  In Sect.~\ref{sec:four}, the procedures of specifying 
the source type and estimating the Einstein radius are explained. The physical quantities of the 
planetary system are estimated in Sect.~\ref{sec:five}, and a summary of results and a conclusion 
are presented in Sect.~\ref{sec:six}.

\section{Data from observations}\label{sec:two}

The source of the lensing event KMT-2021-BLG-0240 lies in a field of the Galactic bulge with 
$({\rm RA}, {\rm decl.})_{\rm J2000} = {\text (17:50:18.55, -30:00:17.89)}$.  The location of 
the source is projected very close to the Galactic center with 
$(l, b) = (-0^\circ\hskip-2pt.387, -1^\circ
\hskip-2pt .426)$, and thus the extinction toward the field, $A_I=3.46$, is considerable. The 
baseline magnitude of the source is $I=21.30$ according to the photometric scale of the Korea 
Microlensing Telescope \citep[KMTNet:][]{Kim2016} survey.

The event was found from the KMTNet survey carried out during the 2021 season.  The survey 
utilizes three telescopes that are distributed in the three sites of the Southern 
Hemisphere for 24-hour monitoring of stars in the bulge field. The sites of the individual 
telescopes are the Siding Spring Observatory, Cerro Tololo Interamerican Observatory, and 
the South African Astronomical Observatory in the three countries of Australia, Chile, and 
South Africa, which are referred to as KMTA, KMTC, and KMTS, respectively.  All telescopes 
are identical with a 1.6~m aperture, and each telescope is mounted by the same wide-field 
camera yielding a 4~deg$^2$ field of view.

\begin{figure}[t]
\includegraphics[width=\columnwidth]{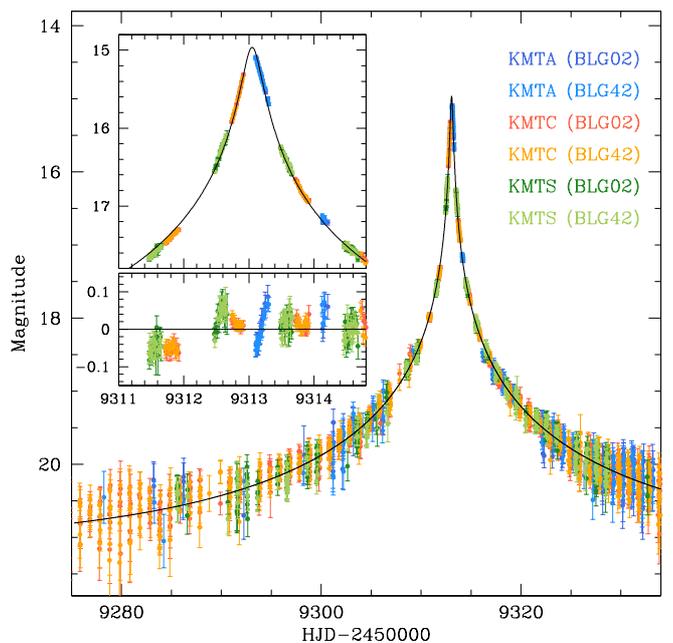}
\caption{
Microlensing light curve of KMT-2021-BLG-0240.  Drawn over the data points is a single-lens 
single-source (1L1S) model curve.  The inset shows the enlargement of the peak region and 
the residuals from the 1L1S model.  The colors of the data points are set to match those of the 
labels designating the telescopes used for observations marked in the legend. 
}
\label{fig:one}
\end{figure}

The lensing event was first found at ${\rm HJD}^\prime\equiv {\rm HJD}-2450000\sim 9307$, 
on 2021 April 5, from the rise of the source flux, which had been constant before the lensing 
magnification.  The event reached its peak at ${\rm HJD}^\prime=9313.03$ (on April 8), and the 
magnification at the peak, $A_{\rm peak}\sim 350$, was very high.  The source is located in 
two of the prime KMTNet fields of BLG02 and BLG42, for which the regions covered by the two 
fields overlap except for the $\sim 15\%$ of the total area that lies in the gaps between chips 
in one of the two fields.  The monitoring cadence for each field was 30~min, and thus the combined 
cadence from the two fields was 15~min. Because the event was observed in two fields of three 
telescopes, the data are composed of 6 sets, which we designate as KMTA (BLG02), KMTA (BLG42), 
KMTC (BLG02), KMTC (BLG42), KMTS (BLG02), and KMTS (BLG42).  From the high-cadence observations 
conducted with the use of the widely separated multiple telescopes, the event was densely covered. 
However, because the event peaked early in the season, when observations could be carried out for 
only $\sim 5$~hr at each observatory, there are approximately 9 hours of gaps over peak, primarily 
on either side of the KMTA observations.  See the enlargement of the peak region presented in the 
inset of Figure~\ref{fig:one}.

\begin{figure}[t]
\includegraphics[width=\columnwidth]{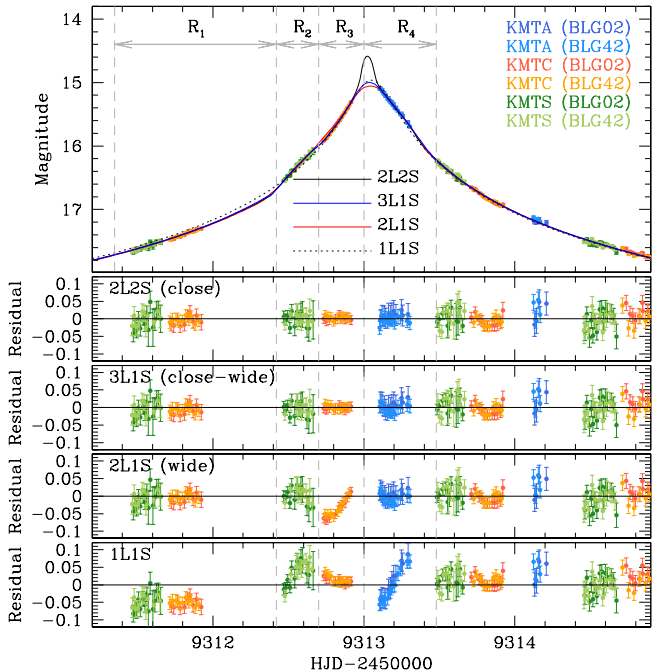}
\caption{
Zoom-in view of the light curve near the peak. The vertical dashed lines denote the four
regions of deviations: 
$9311.35 \leq {\rm HJD}^\prime \leq 9312.42$ ($R_1$ region), 
$9312.42 \leq {\rm HJD}^\prime \leq 9312.70$ ($R_2$ region), 
$9312.70 \leq {\rm HJD}^\prime \leq 9313.00$ ($R_3$ region), and 
$9313.00 \leq {\rm HJD}^\prime \leq 9312.48$ ($R_4$ region). 
The curves over the data represent the 2L2S (close), 3L1S (close-wide), 2L1S (wide), and 1L1S 
models, for which the residuals are shown in the lower panels. The 2L1S model is found by 
modeling the data excluding those in the $R_3$ 
region.
}
\label{fig:two}
\end{figure}

The MOA collaboration \citep{Bond2001} carried out intensive observations of the Galactic bulge 
field during the time gap between KMTC and KMTA data.  Unfortunately, KMT-2021-BLG-0240 lies 
9 minutes of arc west of the boundary of the MOA field, and thus no data from the MOA survey 
are available.

Observations of the event were primarily done in the $I$ band, and a portion of the data were 
obtained in the $V$ band to estimate the source color.  Reductions of images and photometry of 
the source were done employing the KMTNet pipeline that was built based on the pySIS code of 
\citet{Albrow2009} using the difference imaging technique \citep{Alard1998}.  For the estimation 
of the source color, extra photometry was done employing the pyDIA code \citep{Albrow2017} for 
a subset of the data taken from KMTC and KMTS.  We explain in detail about the source type 
specification in Sect.~\ref{sec:four}.  In order to take into consideration the scatter of data 
and to normalize $\chi^2$ value per degree of freedom (dof) for each data set to unity, we 
readjust error bars of data estimated by the automated photometry pipeline, $\sigma_0$, according 
the routine mentioned by \citet{Yee2012}, that is, $\sigma=k(\sigma_0^2+\sigma_{\rm min}^2)^{1/2}$, 
where $\sigma_{\rm min}$ is inserted in the quadrature for the consideration of the data scatter, 
and $k$ is a scaling factor used to make $\chi^2/{\rm dof}=1$.

\section{Anomaly in the lensing light curve}\label{sec:three}

The light curve of the event constructed with the photometric data from the three KMTNet telescopes 
are plotted in Figure~\ref{fig:one}. From a glimpse, it would appear to be that of a normal single-lens 
single-source (1L1S) event with a high magnification.  However, a close look at the light curve 
reveals that it exhibits slight deviations at the 0.1~mag level in the region around the peak.  
A 1L1S model and the residuals are presented in the inset of Figure~\ref{fig:one}.  The deviations 
appear in all three sets of the KMTS, KMTC, and KMTA data taken during the time span $9311.4 \lesssim 
{\rm HJD}^\prime \lesssim 9313.5$.  We note that the gaps among the data sets appear because  the peak 
of the event occurred during the early observing season, at which time the bulge could be observed for 
$\sim 5$~hrs.  We separately show the peak region in Figure~\ref{fig:two} to better illustrate the 
deviations.

\subsection{Binary-lens (2L1S) and source (1L2S) interpretations}\label{sec:three-one}

To explore the origin of the anomaly, we first test a model with two lens components ($M_1$ and $M_2$): 
2L1S model.  We check this model because the anomaly appears near the peak, for which the chance of 
being perturbed by a planetary companion located near the Einstein ring \citep{Griest1998} or a binary 
companion with a very large or a small separation \citep{Han2005} is high.

\begin{table*}[t]
\small
\centering
\caption{Lensing parameters of 3L1S solutions\label{table:one}}
\begin{tabular}{lccccc}
\hline\hline
\multicolumn{1}{c}{Parameter}         &
\multicolumn{1}{c}{close-close}       &
\multicolumn{1}{c}{close-wide}        &
\multicolumn{1}{c}{wide-close}        &
\multicolumn{1}{c}{wide-wide}         \\
\hline
$\chi^2$/dof               &  2663.1/2658              &  2662.8/2658              &   2664.5/2658              &  2663.7/2658               \\
$t_0$ (HJD$^\prime$)       &  $9313.04985 \pm 0.008$   &  $9313.048  \pm  0.004$   &   $9313.040 \pm 0.006 $    &  $9313.042  \pm 0.006 $    \\ 
$u_0$ ($10^{-3}$)          &  $   2.99    \pm 0.25 $   &  $    3.03  \pm  0.18 $   &   $   3.06  \pm 0.20  $    &  $   2.93   \pm 0.18  $    \\
$\te$ (days)               &  $  42.13    \pm 1.83 $   &  $  42.33   \pm  1.61 $   &   $  41.68  \pm 1.94  $    &  $  42.73   \pm 1.89  $    \\
$s_2$                      &  $   0.954   \pm 0.005$   &  $   0.954  \pm  0.005$   &   $   1.064 \pm 0.005 $    &  $   1.061  \pm 0.006 $    \\
$q_2$ (10$^{-3}$)          &  $   0.69    \pm 0.10 $   &  $   0.64   \pm  0.11 $   &   $   0.95  \pm 0.14  $    &  $   0.89   \pm 0.13  $    \\ 
$\alpha$ (rad)             &  $  -0.773   \pm 0.045$   &  $  -0.737  \pm  0.052$   &   $  -0.911 \pm 0.063 $    &  $  -0.907  \pm 0.061 $    \\
$s_3$                      &  $   0.372   \pm 0.090$   &  $   2.721  \pm  0.303$   &   $   0.503 \pm 0.088 $    &  $   2.052  \pm 0.389 $    \\
$q_3$ (10$^{-3}$)          &  $   1.87    \pm 1.29 $   &  $   1.83   \pm  0.50 $   &   $   0.89  \pm 0.81  $    &  $   1.01   \pm 0.70  $    \\ 
$\psi$ (rad)               &  $   1.706   \pm 0.058$   &  $   1.702  \pm  0.057$   &   $   1.781 \pm 0.063 $    &  $   1.805  \pm 0.057 $    \\
$\rho$ (10$^{-3}$)         &  $   2.78    \pm 0.28 $   &  $   2.74   \pm  0.30 $   &   $   2.81  \pm 0.32  $    &  $   2.71   \pm 0.33  $    \\
\hline
\end{tabular}
\tablefoot{ ${\rm HJD}^\prime = {\rm HJD}- 2450000$.  }
\end{table*}

The 2L1S modeling is conducted to search for a solution, defined by a parameter set that 
best describes the observed data.  Among the lensing parameters, three depict the encounter 
between the source and lens: $t_0$, $u_0$, and $\te$, which indicate the epoch and impact 
parameter of the source-lens approach, and time scale of the event, respectively.  Another 
three parameters depict the binary lens system: $s$, $q$, and $\alpha$. These parameters 
denote the $M_1$--$M_2$ separation (in projection and normalized to $\thetae$), the mass 
ratio, and the angle between the source motion and the binary axis, respectively. Besides 
these parameters, we add an extra parameter $\rho$ (normalized source radius), denoting the 
angular source radius $\theta_*$ in units of $\thetae$.  This parameter is included to take 
into consideration finite-source effects that may give rise to the deformation of the anomaly 
during the source crossings over caustics.

The 2L1S lensing parameters were searched for using two approaches, in which the binary parameters
$s$ and $q$ were investigated via a grid approach, and the others were derived via a downhill 
approach.  A Markov Chain Monte Carlo (MCMC) logic was used in the downhill approach.  This grid 
search provided us a $\Delta\chi^2$ map on the $s$--$q$ plane, and the map enabled us to identify 
local minima.  For each local solution, we polished the parameters, including $s$ and $q$, by 
allowing them to vary using the MCMC approach.  This approach of finding lensing solutions is 
useful for identifying degenerate solutions, in which different models result in similar light 
curves.  From the modeling, it was found that a 2L1S interpretation does not yield a model that 
adequately explains the observed anomaly.

Recognizing that the light curve cannot be explained by a 2L1S interpretation, we then checked the 
possibility that the source is composed of binary stars: 1L2S model.  A 1L2S modeling requires four 
extra lensing parameters in addition to those of a 1L1S model.  These extra parameters are 
$t_{0,2}$, $u_{0,2}$, $\rho_2$, and $q_F$, and they represent the epoch and separation of the second 
source ($S_2$) from the lens at the closest approach, the normalized radius of $S_2$, and the flux 
ratio between the primary ($S_1$) and secondary source stars, respectively.  It was found that this 
interpretation did not yield a model describing the anomaly either.  This is because the anomaly 
shows both positive and negative deviations from the 1L1S model, as shown in the residual presented 
in the bottom panel of Figure~\ref{fig:two}, but a 1L2S model can generate only positive deviations.

\begin{figure}[t]
\includegraphics[width=\columnwidth]{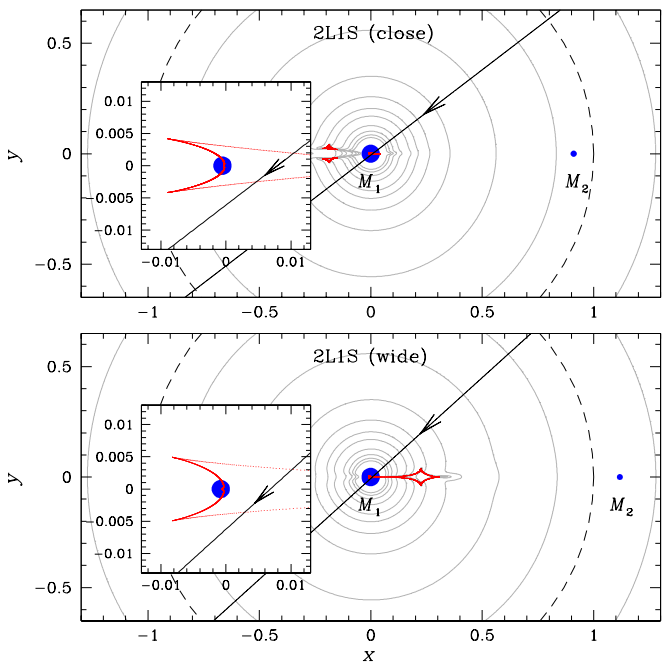}
\caption{
Configurations of the close (upper panel) and wide (lower panel) 2L1S solutions found from the 
modeling conducted by excluding the data in the $R_3$ deviation region, marked in Fig.~\ref{fig:two}.  
In each panel, the line with an arrow indicates the trajectory of the source motion, the concave 
closed curve is the caustic, the dashed circle denotes the Einstein ring, and the two blue dots 
marked by $M_1$ and $M_2$ indicate the positions of the lens components.  The zoom-in view of the 
central magnification region is shown in the inset.  The grey curves encompassing the caustic 
represent the equi-magnification contours.   
}
\label{fig:three}
\end{figure}

We then conduct an additional modeling to check whether a 2L1S model can describe a part of the 
anomaly.  We do this check because if the lens is composed of three masses (3L1S) or the source 
is a binary (2L2S), a 2L1S model can often depict a part of the anomaly, while the rest of the 
anomaly can be described by introducing a tertiary lens component or a companion to the source.  
For this check, we divide the anomaly into four regions:
$9311.35 \leq {\rm HJD}^\prime \leq 9312.42$ ($R_1$ region), 
$9312.42 \leq {\rm HJD}^\prime \leq 9312.70$ ($R_2$ region), 
$9312.70 \leq {\rm HJD}^\prime \leq 9313.00$ ($R_3$ region), and 
$9313.00 \leq {\rm HJD}^\prime \leq 9312.48$ ($R_4$ region). 
The divisions of the regions are marked by dashed vertical lines in Figure~\ref{fig:two}. The 2L1S 
modeling according to this scheme is done for the data excluding those in one of the four deviation 
regions.  From the modeling conducted by excluding the data in the $R_3$ region, we find two models 
that can approximately explain the deviations in the other three regions.  The binary lens parameters 
of these models are 
\begin{equation}
(s, q)\sim 
\begin{cases}
(0.91, 0.6\times 10^{-3}) & ({\rm close}), \\
(1.12, 0.9\times 10^{-3}) & ({\rm wide}).
\end{cases}
\label{eq1}
\end{equation}
The two locals are referred to as ``close'' and ``wide'' solutions because $s<1.0$ and $s>1.0$ for 
the individual solutions.  The model curve and the residual of the close 2L1S solution are shown in 
Figure~\ref{fig:two}.  For the two degenerate solutions, the binary separations approximately follow 
the relation $s_{\rm close}\times s_{\rm wide}\sim 1$, and this suggests that the similarity between 
the two models originates from the close-wide degeneracy first pointed out by \citet{Griest1998} and 
later discussed in detail by \citet{Dominik1999} and \cite{An2005}.  For both solutions, the mass 
ratios between the lens components are less than $10^{-3}$, suggesting that a planetary-mass 
companion accompanies the primary of the lens regardless of the solutions.

Figure~\ref{fig:three} shows the lensing configurations of the close (upper panel) and wide (lower
panel) 2L1S models.  For both solutions, the anomaly was produced by the source crossings over the 
central caustic induced by a planet lying close to the Einstein ring.  Due to severe finite-source 
effects, the light curve during the caustic crossings induce weak deviations instead of usual 
sharp spike features.

\subsection{Triple-lens (3L1S) interpretation}\label{sec:three-two}

The fact that a 2L1S model partially explains the anomaly suggests that there may be a tertiary lens 
component, $M_3$. This is because  an anomaly induced by two companions ($M_2$ and $M_3$), in many 
cases, can be approximately described by the superposition of the 2L1S perturbations, in which the 
pairs of $(M_1, M_2)$ and $(M_1, M_3)$ behave as individual 2L lenses \citep{Bozza1999, Han2001}.  
Under this approximation, then, the residual from the 2L1S model, that is, the deviation in the 
$R_3$ region, may be explained by adding a tertiary lens component. For this check, we conduct a 
3L1S modeling.

The 3L1S modeling was conducted in a similar fashion to the 2L1S modeling. We first found the
parameters related to the tertiary lens component $(s_3, q_3, \psi)$ via a grid approach by keeping 
the other parameters the same as those of the 2L1S solution, and then polished the locals found from 
the preliminary modeling by releasing all lensing parameters as free parameters.  Here $s_3$ and 
$q_3$ denote the separation and mass ratio between $M_1$ and $M_3$, respectively, and $\psi$ indicates 
the orientation angle of $M_3$ as measured from the $M_1$--$M_2$ axis with a center at the position of 
$M_1$. We denote the parameters describing $M_2$ as $(s_2, q_2)$ to distinguish them from those 
describing $M_3$. Because a central caustic can be induced not only by a planet lying near the Einstein 
ring but also by a binary companion with a very large or a small separation \citep{Lee2008}, we set the 
ranges of $s_3$ and $q_3$ wide enough to consider both planetary and binary companions: $[-1.5, 1.5]$ 
for $\log s_3$ and $[-5.0, 0.0]$ for $\log q_3$.

\begin{figure}[t]
\includegraphics[width=\columnwidth]{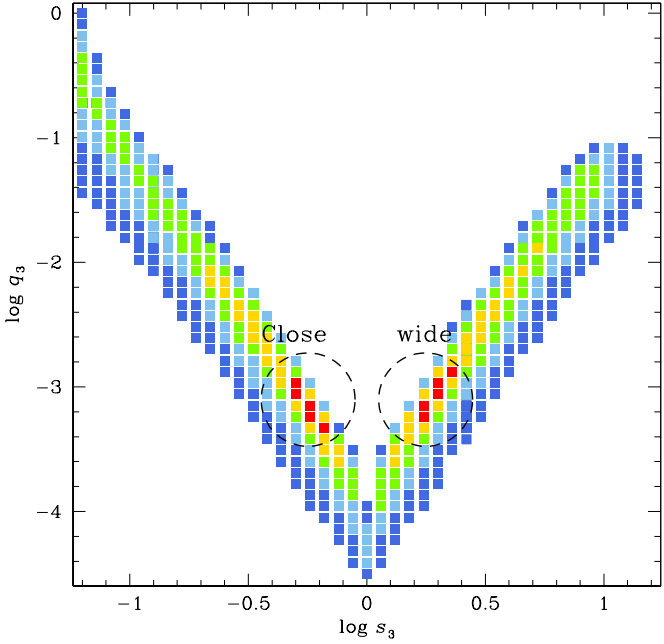}
\caption{
Map of $\Delta\chi^2$ on the $\log~s_3$--$\log~q_3$ plane. The color coding  corresponds to regions 
with $\Delta\chi^2 \leq 1$ (red), $\leq 4$ (yellow), $\leq 9$ (green), $\leq 16$ (cyan), and
 $\leq 25$ (blue).  The two regions enclosed by dashed circles constitute the two locals.
}
\label{fig:four}
\end{figure}

Figure~\ref{fig:four} shows the $\Delta\chi^2$ map on the $\log s_3$--$\log q_3$ plane constructed 
by conducting the grid searches for these parameters with the initial 2L1S parameters adopted from 
those of the close 2L1S solution.  The map shows two distinctive local solutions lying at
\begin{equation}
(\log s_3, \log q_3) \sim 
\begin{cases}
(-0.3, -3),    & \text{(close)},\\
(+0.3, -3),    & \text{(wide)}.
\end{cases}
\label{eq2}
\end{equation}
The two solutions result in similar model curves caused by the close--wide degeneracy in $s_3$, 
and thus we refer them to as `close' and `wide' solutions.  We find another two solutions obtained 
from the modeling with the initial parameters of the wide 2L1S solution, and thus there exist 4 
solutions in total.  We designate the individual solutions as close-close ($s_2<1.0$ and $s_3<1.0$), 
close-wide ($s_2<1.0$ and $s_3>1.0$), wide-close ($s_2>1.0$ and $s_3<1.0$), and wide-wide ($s_2>1.0$ 
and $s_3>1.0$) solutions.  The lensing parameters of the four 3L1S models obtained under the assumption 
of a rectilinear relative lens-source motion (standard model) are presented in Table~\ref{table:one} 
together with the values of $\chi^2$/dof.  It was found that the degeneracies among the solutions are 
severe with $\Delta\chi^2 < 1.7$.  We note that it is difficult to choose a correct model based on the 
argument on the dynamical stability of the lens system first because both the companions are planets 
for which their masses are too small to affect the dynamics of the system unlike triple stellar systems 
\citep{Toonen2016}, and second because the measured separations are not intrinsic values but projected 
ones.

It is found that the 3L1S models can explain all the features of the anomaly. The model curve of 
the best-fit 3L1S model (close-wide model) and its residuals are presented in Figure~\ref{fig:two}.  
We note that the other 3L1S solutions yield  similar models to the presented one. The estimated mass 
ratios between the $M_1$--$M_2$ pair are in the range of $q_2\sim [0.6$--1.0]$\times 10^{-3}$, and 
those between the $M_1$--$M_3$ pair are in the range of $q_3\sim [0.9$--1.9]$\times 10^{-3}$.  These 
mass ratios roughly correspond to the ratio between the Jupiter and the Sun of the Solar system.  
According to the 3L1S interpretation, then, the lens is a planetary system possessing two giant 
planets.   If this interpretation is correct, the lens of the event is the sixth case of multiple 
planetary system found by microlensing, following OGLE-2006-BLG-109 \citep{Gaudi2008, Bennett2010}, 
OGLE-2012-BLG-0026 \citep{Han2013, Beaulieu2016}, OGLE-2018-BLG-1011 \citep{Han2019}, OGLE-2019-BLG-0468 
\citep{Han2022c}, and KMT-2021-BLG-1077 \citep{Han2022a}.

\begin{figure}[t]
\includegraphics[width=\columnwidth]{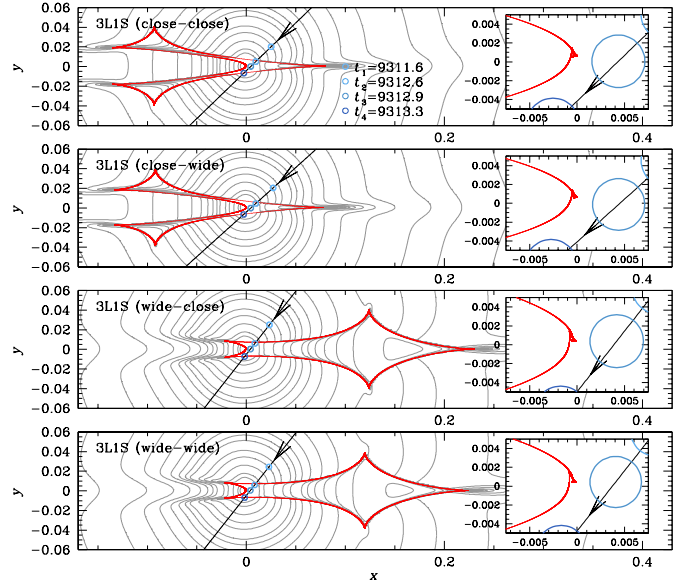}
\caption{
Lensing configurations of the four 3L1S solutions: close-close, close-wide, wide-close, and wide-wide 
solutions.  The right inset in each panel displays the magnified view around the planet host.  The 
four empty cyan circles on the source trajectory (labeled as $t_1$, $t_2$, $t_3$, and $t_4$) denote 
the source positions corresponding to the deviations in the $R_1$, $R_2$, $R_3$, and $R_4$ regions. 
The circle size is scaled to $\thetae$ corresponding to the total mass of the lens.  The grey curves 
represent the equi-magnification contours.  
}
\label{fig:five}
\end{figure}

Figure~\ref{fig:five} shows the lensing configurations corresponding to the four 3L1S solutions.  
For both close-xx and wide-xx solutions, the caustics appear to be similar to those of the 
corresponding 2L1S solutions presented in Figure~\ref{fig:three}, but they differ from the 2L1S 
caustics in the region around the planet host.  See the zoom-in view of the central region shown in 
the right inset of each panel.  In this region, there exists a tiny caustic generated by the tertiary 
lens component. The four empty cyan circles (labeled as $t_1$, $t_2$, $t_3$, and $t_4$) on the 
source trajectory denote the source positions corresponding to the deviations in the $R_1$, $R_2$, 
$R_3$, and $R_4$ regions, respectively, marked in Figure~\ref{fig:two}.  It shows that the source 
passed the region of positive deviations extending from a cusp of the tiny caustic induced by $M_3$, 
and this explains the $R_3$ region deviation that could not be accounted for by the two-mass lens 
model.

We checked higher-order effects causing nonrectilinear relative motion between the source and lens.
Two major effects cause deviations from the rectilinear motion: microlens-parallax and lens-orbital 
effects. The former effect arises due to the Earth's (observer's) orbital motion around the Sun 
\citep{Gould1992}, and the latter effect arises due to the orbital motion of the lens \citep{Dominik1998}.  
We found that specifying the lens orbital motion was difficult because the anomaly lasted for a very 
short period of time, and thus only the parallax effect was considered in the modeling.  The extra 
parameters required to define the microlens-parallax effect are the two components ($\pien$ for 
the north direction and $\piee$ for the east direction) of the microlens-lens parallax vector 
$\pivec_{\rm E}=(\pi_{\rm rel}/\thetae)(\muvec/\mu)$, where $\pi_{\rm rel}={\rm AU}
(D_{\rm L}^{-1}-D_{\rm S}^{-1})$ represents the relative lens-source parallax, $D_{\rm L}$ and 
$D_{\rm S}$ represent the distances to the lens and source, respectively, and $\muvec$ is the relative 
lens-source proper motion.  We investigated whether microlens parallax could be constrained, but it 
was found that the small parallax signal was not consistent between observatories and therefore was 
most likely due to low-level systematics, which often dominate the parallax signals for events with 
faint source stars.

\begin{table}[t]
\small
\caption{Lensing parameters of 2L2S model\label{table:two}}
\begin{tabular*}{\columnwidth}{@{\extracolsep{\fill}}lcc}
\hline\hline
\multicolumn{1}{c}{Parameter}           &
\multicolumn{1}{c}{Close}               &
\multicolumn{1}{c}{Wide}                \\
\hline
$\chi^2$/dof             &   2652.0/2658             &  2652.5/2658               \\
$t_{0,1}$ (HJD$^\prime$) &    $9313.054 \pm  0.005$  &   $9313.050 \pm  0.004$    \\
$u_{0,1}$ ($10^{-3}$)    &    $   4.54  \pm  0.33 $  &   $   4.14  \pm  0.28 $    \\
$t_{0,2}$ (HJD$^\prime$) &    $9313.019 \pm  0.006$  &   $9313.019 \pm  0.007$    \\
$u_{0,2}$ ($10^{-3}$)    &    $   0.08  \pm  0.19 $  &   $   0.02  \pm  0.24 $    \\
$\te$ (days)             &    $  40.24  \pm  1.58 $  &   $  42.04  \pm  1.64 $    \\
$s$                      &    $   0.958 \pm  0.006$  &   $   1.059 \pm  0.006$    \\
$q$ ($10^{-3}$)          &    $   0.36  \pm  0.09 $  &   $   0.41  \pm  0.11 $    \\
$\alpha$ (rad)           &    $  -0.544 \pm  0.044$  &   $  -0.579 \pm  0.051$    \\
$\rho_1$ ($10^{-3}$)     &    $   2.01  \pm  0.35 $  &   $   1.98  \pm  0.30 $    \\
$\rho_2$ ($10^{-3}$)     &    $   1.07  \pm  0.20 $  &   $   1.05  \pm  0.20 $    \\
$q_{F,I}$                &    $   0.18  \pm  0.05 $  &   $   0.14  \pm  0.03 $    \\
\hline
\end{tabular*}
\end{table}

\subsection{Binary-lens binary-source (2L2S) interpretation}\label{sec:three-three}

It is known that a 3L1S model can occasionally be degenerate with a 2L2S model, in which both the 
lens and source are binaries, as illustrated in the cases of the lensing events KMT-2019-BLG-1953 
\citep{Han2020-1953} and OGLE-2018-BLG-0532 \citep{Ryu2020}.  For the investigation of this degeneracy, 
we additionally conducted a 2L2S modeling of the event.  In this modeling, we started with the lensing 
parameters of the 2L1S model explaining the anomaly features except the $R_3$ region, and checked a 
possible trajectory of $S_2$ explaining the anomaly in the $R_3$ region.

From the 2L2S modeling, we found two solutions that well explain all of the anomaly features.  The 
individual solutions were found based on the close and wide 2L1S solutions, and thus we designate 
them as ``close'' and ``wide'' solutions.  The close model yields a slightly better fit to the data, 
but the difference is very small with $\Delta\chi^2=0.5$.  The model curve and residual of the close 
2L2S solution are shown in Figure~\ref{fig:two}, and the lensing parameters of both solutions are 
listed in Table~\ref{table:two}.  In the table, we use the notations $(t_{0,1}, u_{0,1}$, $\rho_1)$ 
to denote the lensing parameters related to the primary source $S_1$.  The 2L2S model curve differs 
from that of the 3L1S model in the time domain $9312.9\lesssim {\rm HJD}^\prime \lesssim 9313.1$, 
but this region corresponds to the gap between the KMTC and KMTA data sets.

\begin{figure}[t]
\includegraphics[width=\columnwidth]{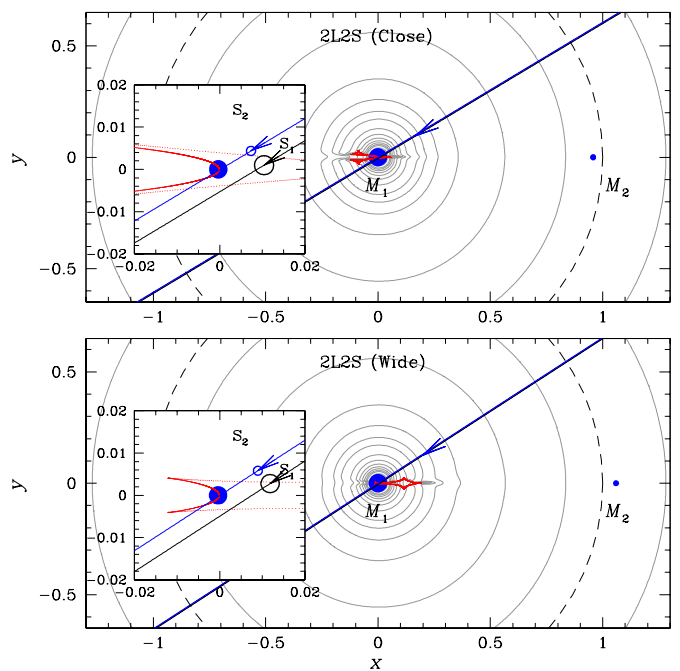}
\caption{
Lensing configurations of the two 2L2S models.  Notations are same as those in Fig.~\ref{fig:three}, 
except that there are two source trajectories, which are marked by $S_1$ for the primary source 
and $S_2$ for the secondary source.  The two source trajectories in the main panel are difficult 
to distinguish due to the closeness of the trajectories, and thus we present the enlargement 
of the central region in the inset.  The trajectories of the primary ($S_1$) and secondary ($S_2$) 
source stars are marked in black and blue lines, respectively. The empty circles on the individual 
source trajectories indicate the source sizes estimated from the 2L2S modeling. 
}
\label{fig:six}
\end{figure}

Figure~\ref{fig:six} shows the lens system configurations of the close (upper panel) and wide (lower 
panel) 2L2S models.  They are similar to the configurations of the corresponding 2L1S models, 
presented in Figure~\ref{fig:three}, except that there are two source trajectories, which are marked 
as $S_1$ for the primary source and $S_2$ for the secondary source.  It is estimated that the ratios 
of the $I$-band flux from the second source to the flux from the primary are $q_{F,I}=0.18$ and 0.14 
according to the close and wide solutions, respectively.  In both cases, the second source, which lay
very close to the primary source, moved in parallel with the primary source, and crossed the caustics 
almost at the same times of the primary source caustic crossings, and thus the caustic crossings of 
$S_2$ did not exhibit extra caustic-crossing features.  However, the second source additionally passed 
over the tiny cusp lying very close to the primary lens, and this gives rise to an extra anomaly feature
that explains the deviation in the $R_3$ region.

It is found that the 2L2S models yield better fits than the 3L1S models by $\Delta\chi^2=[10.8-12.5]$.
If the 2L2S interpretation is correct, KMT-2021-BLG-0240 is the sixth lensing event for which both the 
lens and source are binaries, following MOA-2010-BLG-117 \citep{Bennett2018}, OGLE-2016-BLG-1003 
\citep{Jung2017}, KMT-2018-BLG-1743 \citep{Han2021a}, KMT-2019-BLG-0797 \citep{Han2021b}, and 
KMT-2021-BLG-1898 \citep{Han2022b}.  Except for OGLE-2016-BLG-1003, the lenses of the other five 
events, including the event analyzed in this work, are planetary systems, indicating that planetary 
signals detectable through the high-magnification channel are prone to be affected by close companions 
to source stars.

\subsection{3L1S versus 2L2S interpretations}\label{sec:three-four}

According to the 2L2S model, the closeness of the source stars should induce large effects on the 
anomaly caused by the orbital motion of the binary source unless they are seen in extreme projection.  
Under the assumption that the binary source orbit is seen face on, the separation between $S_1$ and 
$S_2$ is $\Delta u = (\Delta\tau^2+ \Delta u_0^2)^{1/2}=4.54\times 10^{-3}$, where $\Delta\tau = 
(t_{0,1}-t_{0,2})/\te$ and $\Delta u_0 = u_{0,1}-u_{0,2}$.  As will be discussed in Sect.~\ref{sec:four}, 
the angular Einstein radius and the relative lens-source proper motion estimated from the 2L2L 
solution are $\thetae\sim 0.44$~mas and  $\mu_{\rm rel}\sim 3.9~{\rm mas}~{\rm yr}^{-1}$, respectively.  
Then, the projected physical separation between the two source stars is $\Delta a_\perp= D_{\rm S} 
\thetae\Delta u \sim 0.016$~AU.  Assuming that $M_{S_1} = 1~M_\odot$ and $M_{S_2}=0.6~M_\odot$, this 
creates an orbital period $P$, an internal velocity $v_{\rm int}$, and an internal proper motion $\mu_{\rm int}$ 
of $P = \{ (a_\perp/{\rm AU})^3 /[(M_{S_1}+M_{S_2})/M_\odot] \}^{1/2}\sim 1.6\times 10^{-3}~{\rm yrs}= 
0.58~{\rm days}$, $v_{\rm int} = (a/P)v_\oplus \sim 300~{\rm km}~{\rm s}^{-1}$, and $\mu_{\rm int} = 
v_{\rm int}/\ds \sim 7.9~{\rm mas}~{\rm yr}^{-1}$, respectively.  Here $(M_{S,1}, M_{\rm S,2})$ denote 
the masses of the source stars, and $v_\oplus=30~{\rm km}~{\rm s}^{-1}$ is the orbital speed of Earth 
around the Sun.  Then, the internal motion induced by the source orbital motion is twice faster than 
the relative lens-source proper motion estimated from the normalized source radius $\rho$, that is, 
$\mu_{\rm rel}\sim 3.9~{\rm mas}~{\rm yr}^{-1}$.  This implies that the normalized source radius 
estimated under static and orbiting source can be different by a factor 2.  Moreover, the positional 
change of the primary source during $P/2=0.29~{\rm days}$ in units of $\thetae$ is $\sim 2(M_{S_2}/M_{S_1}) 
\Delta u \sim 3.4\times 10^{-3}$, which is similar to $u_0$.  Considering that the effective time scale 
of the event is $t_{\rm eff}=u_0\te\sim 0.18~{\rm days}$, the light curve would show extremely violent 
(factor 2) changes by the orbital motion of the source during $\sim [-t_{\rm eff},+t_{\rm eff}]$, but 
no such changes were observed.

The 2L2S solution is further disfavored by the absence of ellipsoidal variations.  If the binary 
source stars are closely separated as measured by the 2L2S solutions, the source would be distorted 
by tides and the light curve would therefore show ellipsoidal variations, but no such variations 
are seen in the light curve.  These problems can be avoided by the assumption that the binary is 
seen in projection or the orbital plane is perpendicular to the direction of lens-source.  This 
argument requires that the proper motion induced by the binary source orbit to be at least 6 times 
smaller than the relative lens-source proper motion.  This can be achieved with a projection factor 
of 200 (probability: $\sim 1/80,000$), or for example, by projection factor of 5 but orbital 
alignment within about $10^\circ$ of perpendicular (probability: $\sim 1/450$).

In order to have confidence in the very low-probability binary-source configurations, one would need  
high statistical confidence.  However, the signal is rather weak for such an analysis.  Furthermore, 
there is an alternative 3L1S solution that is disfavored by only $\Delta\chi^2 \sim 10$, which is 
small given the quality of the data.  Therefore, we conclude that the 3L1S and 2L2S solutions cannot 
be distinguished with the available data, and either could be correct.  However, we note that the 
detection of one planet, $M_2$ for the 3L1S model, is solid regardless of the solutions.  In the 
following analysis, we estimate the physical lens parameters for both interpretations of the event.

Because of the possible importance of the source orbital motion together with no clear features of 
sharp caustic crossings, the observed anomaly could be, in principle, entirely due to ``xallarap'' 
effects, that is, the orbital effect induced by a source companion whose own luminosity contributes 
negligibly to the light curve of a single-lens event.  Motivated by this consideration, we check 
whether the anomaly could be due to xallarap effects alone.  The modeling considering the xallarap 
effect requires 5 additional parameters in addition to the 1L1S parameters: $\xien$, $\xiee$, $P$, 
$\phi$, and $i$ \citep{Dong2009}.  The parameters $(\xien, \xiee)$ denote the north and east components 
of the xallarap vector $\xivec_{\rm E}$, $P$ is the orbital period, and $(\phi, i)$ denote the phase 
and inclination angles of the orbit, respectively.  The magnitude of $\xivec_{\rm E}$ is related to 
the semi-major axis, $a$, of the source orbit by $\xi_{\rm E}=a_{S}/(\ds\thetae)$, where $a_{\rm S}=
a M_{S,2}/(M_{\rm S,1}+M_{S,2})$.  Considering the short duration of the anomaly, we test xallarap 
models with orbital periods within the range of $0.1\leq  P/{\rm days}\leq 10$.  We find that the 
fits of the xallarap models are worse than the 3L1S and the static 2L2S models by $\Delta\chi^2 
\sim 42$ and $\sim 53$, respectively, and thus we conclude that the anomaly cannot be attributed 
to the xallarap effect.

\section{Source star and angular Einstein radius}\label{sec:four}

In this section, we estimate the extra observable of $\thetae$ according to the 3L1S and 2L2S 
solutions.  For both solutions, $\thetae$ measurement is possible because the source star crossed 
the caustic and the light curve around these epochs was impacted by the effect of a finite source 
size.  The measured normalized source radius allows one to estimate the Einstein radius 
\citep{Gould1994, Witt1994, Nemiroff1994} as
\begin{equation}
\thetae = {\theta_* \over \rho}.
\label{eq3}
\end{equation}

For the measurement of $\thetae$, one needs to measure $\theta_*$, which is inferred from the color 
and brightness of the source.  For the estimation of the source color, it is required to measure the 
source magnitudes in two passbands, $V$ and $I$ in our case, from the regression of the photometric 
data to the model of the lensing light curve.  For KMT-2021-BLG-0240, we could measure the $I$-band 
magnitude precisely, but it was difficult to measure a reliable $V$-band magnitude because of the 
poor quality of the $V$-band data caused by the heavy extinction toward the field.  Due to this 
difficulty, we interpolate the source color from the main-sequence branch of stars in the color-magnitude 
diagram (CMD) built from the observations using the {\it Hubble Space Telescope} ({\it HST}) 
\citep{Holtzman1998} based on the measured $I$-band magnitude.

\begin{figure}[t]
\includegraphics[width=\columnwidth]{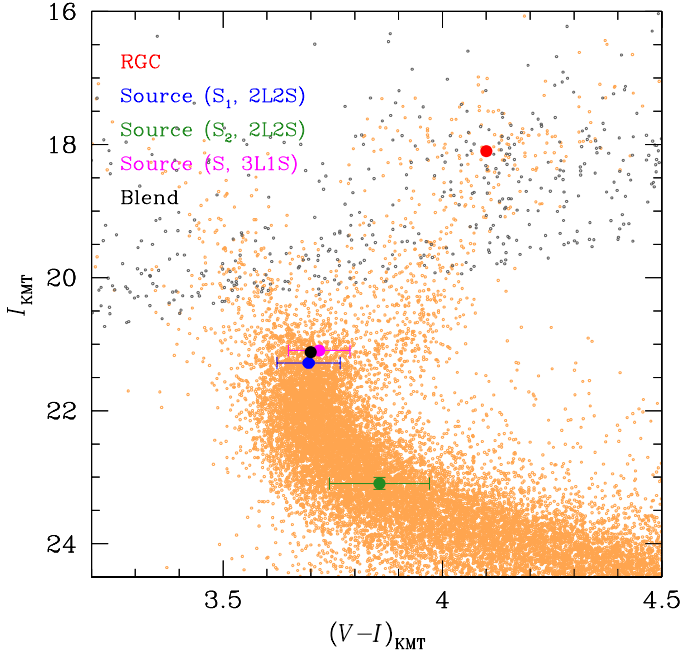}
\caption{
Color-magnitude diagram (CMD) constructed from the combination of the {\it Hubble Space Telescope}
and ground-based (KMTNet) observations.  The small filled dot marked in magenta indicates the 
source position according to the 3L1S solution, and the blue and green dots represent the positions 
of the primary and secondary source stars estimated from the 2L2S solution, respectively.  The red 
dot denotes the centroid of red giant clump (RGC).  
}
\label{fig:seven}
\end{figure}

Figure~\ref{fig:seven} shows the location of the source star in the CMD constructed from the 
combination of {\it HST} and ground-based observations.  The ground-based CMD was built using 
the pyDIA photometry of the KMTC data, and the {\it HST} and KMTC CMDs were aligned utilizing 
the centroids of the red giant clump (RGC) on the individual CMDs.  In the CMD, the small filled 
dot marked in magenta indicates the source position according to the 3L1S solution, and the blue 
and green dots represent the positions of the primary and secondary source stars estimated from 
the 2L2S solution, respectively.  In the case of the 2L2S solution, the $I$-band magnitudes of 
the two source stars were estimated from the combined flux $F_{{\rm S},I}$, which was measured 
from the modeling, by 
\begin{equation}
\eqalign{
F_{S_1,I} = \left( {1 \over 1+q_{F,I}} \right)F_{S,I};\qquad 
F_{S_2,I} = \left( {q_{F,I} \over 1+q_{F,I}} \right)F_{S,I}, 
}
\label{eq4}
\end{equation}
where $q_{F,I}$ is the flux ratio between the two source stars, and $F_{S_1,I}$ and $F_{S_2,I}$ 
indicate flux values of $S_1$ and $S_2$, respectively.  
The measured values of the instrumental (uncalibrated) color and magnitude are
\begin{equation}
(V-I, I)_{S,{\rm 3L1S}} = (3.72 \pm 0.07, 21.10 \pm 0.01), 
\label{eq5}
\end{equation}
for the 3L1S solution, 
\begin{equation}
\eqalign{
 & (V-I, I)_{S_1, {\rm 2L2S}} = (3.70 \pm 0.07, 21.28 \pm 0.02), \cr
 & (V-I, I)_{S_2, {\rm 2L2S}} = (3.86 \pm 0.11, 23.10 \pm 0.09)  \cr
}
\label{eq6}
\end{equation}
for the 2L2S solution, and
\begin{equation}
(V-I, I)_{\rm RGC} = (4.10, 18.10)
\label{eq7}
\end{equation}
for the RGC centroid commonly for the both solutions.  Also marked in the CMD is the location 
of a blend (black filled dot).  It shows that the observed flux is affected by blended light 
that is almost as bright as the source.

Calibration of the source color and magnitude was done by applying the method 
of \citet{Yoo2004}, which utilizes the RGC centroid with its known dereddened color and magnitude, 
$(V-I, I)_{{\rm RGC},0}$.  Following this method, the dereddened values of the source were estimated 
using the known values of the RGC centroid, $(V-I)_{{\rm RGC},0}=(1.06, 14.45)$ \citep{Bensby2013, 
Nataf2013}, and the offsets between the source and RGC centroid, $\Delta (V-I, I)= (V-I, I)_{S}-
(V-I, I)_{\rm RGC}$, by $(V-I, I)_{S,0} = (V-I, I)_{{\rm RGC},0} + \Delta (V-I, I)$.  
The dereddened values of the color and magnitude for the source estimated from this calibration 
process are 
\begin{equation}
(V-I, I)_{S,0,{\rm 3L1S}} = (0.68 \pm 0.07, 17.44 \pm 0.01),
\label{eq8}
\end{equation}
for the 3L1S solution, and 
\begin{equation}
\eqalign{
 & (V-I, I)_{S_1,0,{\rm 2L2S}} = (0.66 \pm 0.07, 17.63 \pm 0.02), \cr
 & (V-I, I)_{S_2,0,{\rm 2L2S}} = (0.82 \pm 0.11, 19.44 \pm 0.09), \cr
}
\label{eq9}
\end{equation}
for the 2L2S solution.
According to the 3L1S solution, the source is a subgiant or a turnoff star of an early G spectral 
type.  According to the 2L2S solution, the secondary source is a late G dwarf and the spectral 
type of the primary source is similar to that of the 3L1S solution but slightly bluer and fainter.

With the measured normalized source radius, we estimated the angular Einstein radius using the 
relation in Equation~(\ref{eq3}).  For the 2L2S solution, the Einstein radius can be estimated 
either by $\thetae=\theta_{*,S_1}/\rho_1$ or $\thetae=\theta_{*,S_2}/\rho_2$, where $\theta_{*,S_1}$ 
and $\theta_{*,S_2}$ represent the angular source radii of $S_1$ and $S_2$, respectively.  We choose 
to estimate $\thetae$ using the former relation because both the stellar type and normalized source 
radius of the primary source  are better constrained than those of the secondary source.
For the estimation 
of the source radius, we first converted $V-I$ color into $V-K$ color using  the \citet{Bessell1988} 
relation, and then assessed $\theta_*$ with the application of the \citet{Kervella2004} relation 
between $V-K$ and $\theta_*$.  The source radii assessed from this procedure are
\begin{equation}
\eqalign{
\theta_{*,S, {\rm 3L1S}}   = 0.98 \pm  0.10~\mu{\rm as},\ \cr
\theta_{*,S_1, {\rm 2L2S}} = 0.88 \pm  0.09~\mu{\rm as}, \cr
}
\label{eq10}
\end{equation}
for the 3L1S and 2L2S solutions, respectively.  From the relation in Equation~(\ref{eq3}), we estimate 
the Einstein radius as 
\begin{equation}
\thetae = 
\begin{cases}
\theta_{*,S,{\rm 3L1S}}/\rho     = 0.35 \pm  0.05~{\rm mas}     & {\rm for\ 3L1S}, \\
\theta_{*,S_1,{\rm 2L2S}}/\rho_1 = 0.44 \pm  0.09~{\rm mas} & {\rm for\ 2L2S}.
\end{cases}
\label{eq11}
\end{equation}
The relative proper motion between the lens and source is estimated by $\mu_{\rm rel}=\thetae/\te$, 
and the values resulting from the $\thetae$ values in Equation~(\ref{eq11}) are
\begin{equation}
\mu_{\rm rel} = 
\begin{cases}
3.06 \pm  0.43~{\rm mas\ yr}^{-1} & {\rm for\ 3L1S}, \\
3.91 \pm  0.79~{\rm mas\ yr}^{-1} & {\rm for\ 2L2S}.
\end{cases}
\label{eq12}
\end{equation}

\section{Physical parameters of planetary system}\label{sec:five}

The physical parameters of a lens system can be constrained by the lensing observables including  
$\te$, $\thetae$, and $\pie$.  Measurements of all these observables enable one to uniquely 
determine the physical parameters as
\begin{equation}
M = {\thetae \over \kappa\pie  };\qquad \dl = { {\rm AU} \over \pie\thetae+\pi_{\rm S}}, 
\label{eq13}
\end{equation}
where $\kappa=4G/(c^2{\rm AU})$, and 
$\pi_{\rm S}={\rm AU}/D_{\rm S}$ is the parallax of the source.  For KMT-2021-BLG-0240, 
the value of $\pie$ cannot be securely measured, although the $\te$ and $\thetae$ observables 
are precisely measured.  Therefore, we estimate $M$ and $\dl$ from a Bayesian analysis utilizing 
a Galactic model based on the constraints given by the measured observables.

In the Bayesian analysis, we generate many artificial lensing lensing events by performing a 
Monte Carlo simulation, in which the locations and motions of sources and lenses and the masses 
of lenses are assigned based on a Galactic model.  In the simulation, we used the \citet{Jung2021} 
Galactic model, which was constructed using the models of \citet{Robin2003} and \citet{Han2003} 
for the physical distributions of disk and bulge objects, respectively, the models of 
\citet{Jung2021} and \citet{Han1995} for the dynamical distributions for disk and bulge objects, 
respectively, and the model of \citet{Jung2018} for the common mass function of disk and bulge 
lenses. The Bayesian posteriors are constructed for the simulated events having observables lying 
in the ranges of the measured values.  As mentioned, the value of the microlens parallax cannot 
be reliably determined for either the 3L1S or the 2L2S solutions, and thus we use the constraints 
of $\te$ and $\thetae$ in the Bayesian analysis.

\begin{figure}[t]
\includegraphics[width=\columnwidth]{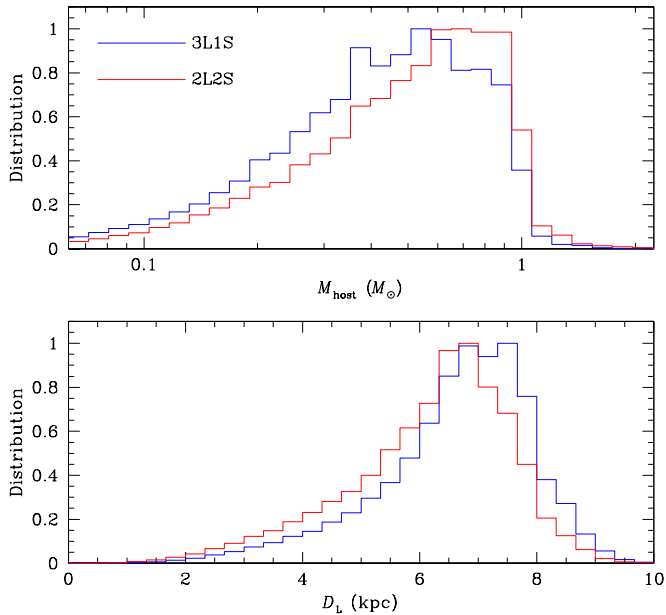}
\caption{
Bayesian posteriors of the host mass and the distance to the planetary system.  The curves 
drawn in blue and red represent the distributions estimated from the 3L1S and 2L2S solutions, 
respectively.
}
\label{fig:eight}
\end{figure}

Figure~\ref{fig:eight} shows the Bayesian posteriors for the mass of the planet host (upper panel) 
and distance to the planetary system (lower panel).  The distributions for the 3L1S and 2L2S solutions 
are marked in blue and red, respectively.  Because of the larger $\thetae$ value, the mass estimated 
from the 2L2S solution is slightly larger than the mass assessed from the 3L1S solution.  For the same 
reason, the distance to the lens for the 2L2S solution is slightly smaller than the distance for the 
3L1S solution.

In Table~\ref{table:three}, we list the physical parameters of the planetary system estimated based 
on the 3L1S and 2L2S solutions.  Common parameters for both solutions include the masses of the host, 
$M_1$, and the confirmed planet, $M_2$, distance, $\dl$, and projected separation between $M_1$ and 
$M_2$, $a_\perp$($M_1$--$M_2$).  For the 3L1S solution, additional parameters of the mass of the 
not-confirmed second planet, $M_3$, and the projected separation between $M_1$ and $M_3$, 
$a_\perp$($M_1$--$M_3$), are listed.  According to the 3L1S solution, the lens is a planetary system 
with two sub-Jovian-mass planets, in which the planets have masses of 0.32--0.47~$M_{\rm J}$ and 
0.44--0.93~$M_{\rm J}$ orbiting an M dwarf host.  According to the 2L2S solution, the lens 
is a planetary system in which a single planet with a mass of $\sim 0.21~M_{\rm J}$ orbits a late 
K-dwarf host.  The distance to the planetary system varies depending on the solution: $\sim 7.0$~kpc 
for the 3L1S solution and $\sim 6.6$~kpc for the 3L1S solution.

Although the two solutions cannot be distinguished based on present data, the degeneracy can in 
principle be broken by future radial velocity (RV) observations. Such observations would find 
the two $q\sim 10^{-3}$ planets if the 3L1S solution is correct, but only one such planet if 
the 2L2S solution is correct.  The first step would be to resolve the host, which will be 
possible at first adaptive-optics (AO) light on next-generation (that is, 30~m class) telescopes. 
For example, by 2030, the source and lens will be separated by the order of 30~mas, according to 
Equation~(\ref{eq12}). Scaling from the experience with current 8~m to 10~m telescopes, it should 
be possible to determine the mass, distance and infrared brightness of the host.  Based on these 
results, it can be determined whether RV observations are feasible with these next-generation 
telescopes or will have to wait for further generations of telescopes, that is, of 100~m diameter.  
Regardless, the wait time is likely to be less than the 57 years between the year that 
\citet{Einstein1936} argued that there would be no great chance of observing this phenomenon and 
that of the first detection of a microlensing event \citep{Alcock1993, Udalski1993}.

\begin{table}[t]
\small
\caption{Physical lens parameters\label{table:three}}
\begin{tabular*}{\columnwidth}{@{\extracolsep{\fill}}llll}
\hline\hline
\multicolumn{2}{c}{Quantity}      &
\multicolumn{1}{l}{3L1S}          &
\multicolumn{1}{l}{2L2S}          \\
\hline
$M_1$ ($M_\odot$)               &               &  $0.47^{+0.33}_{-0.24}$             &  $0.56^{+0.32}_{-0.30}$     \\  [0.7ex]
$M_2$ ($M_{\rm J}$)             &               &  0.32 -- 0.47                       &  $0.21^{+0.12}_{-0.11}$     \\  [0.7ex]
$M_3$ ($M_{\rm J}$)             &               &  0.44 -- 0.93                       &  --                         \\  [0.7ex]
$\dl$ (kpc)                     &               &  $7.0^{+1.0}_{-1.5}  $              &  $6.6^{+1.0}_{-1.7}   $     \\  [0.7ex]

$a_\perp$ ($M_1$--$M_2$) (AU)   & close         &  $2.5^{+2.8}_{-1.9}  $              &  $2.8^{+3.3}_{-2.1}   $     \\  [0.7ex]
                                & wide          &  $2.7^{+3.1}_{-2.2}  $              &  $3.1^{+3.6}_{-2.3}   $     \\  [0.7ex]     

$a_\perp$ ($M_1$--$M_3$) (AU)   & close-close   &  $1.0^{+1.1}_{-0.8}  $              &  --                         \\  [0.7ex]
                                & close-wide    &  $7.0^{+8.0}_{-5.5}  $              &  --                         \\  [0.7ex]     
                                & wide-close    &  $1.3^{+1.5}_{-1.0}  $              &  --                         \\  [0.7ex]     
                                & wide-wide     &  $5.3^{+6.0}_{-4.2}  $              &  --                         \\  [0.7ex]     
\hline                            
\end{tabular*}
\tablefoot{
The arrows in the third column indicate that the values are same as those in the second column.}
\end{table}

\section{Summary and conclusion}\label{sec:six}

We investigated the lensing event KMT-2021-BLG-0240, for which the light curve was densely 
and continuously covered from the high-cadence observations using the globally distributed 
three telescopes of the KMTNet survey conducted in the 2021 season.  The light curve from a 
glimpse appeared to be that of a regular lensing event produced by a single mass magnifying 
a single source star, but a close look revealed an anomaly with complex features at the 0.1~mag 
level lasted for $\sim 2$~days in the region around the peak.

It was found that the anomaly could not be explained with either a 2L1S or a 1L2S model, which 
are the most common causes of microlensing anomalies.  However, we found that a 2L1S model could 
describe a part of the anomaly, suggesting the possibility that the anomaly might be deformed by 
a tertiary lens component or a close companion to the source.  From the additional modeling, we 
found that all the features of the anomaly could be explained with either a 3L1S model or a 2L2S 
model.  In the sense of the goodness of the fit, the 2L2S interpretation was slightly preferred 
over the 3L1S interpretation.  However, the 2L2S interpretation was less favored due to the 
absence of signatures induced by the source orbital motion and ellipsoidal variations.  We, 
therefore, conclude that the two interpretations could not be distinguished with the available 
data, and either could be correct.

According to the 3L1S solution, the lens is a planetary system with two sub-Jovian-mass planets, 
in which the planets have masses of 0.32--0.47~$M_{\rm J}$ and 0.44--0.93~$M_{\rm J}$ and they 
orbit an M dwarf host.  According to the 2L2S solution, on the other hand, the lens is a single 
planet system with a $\sim 0.21~M_{\rm J}$ planet orbiting a late K-dwarf host, and the source 
is a binary composed of a primary of a subgiant or a turnoff star and a secondary of a late G dwarf.  
The distance to the planetary system varies depending on the solution: $\sim 7.0$~kpc for the 3L1S 
solution and $\sim 6.6$~kpc for the 2L2S solution.

\begin{acknowledgements}
Work by C.H. was supported by the grants  of National Research Foundation of Korea 
(2020R1A4A2002885 and 2019R1A2C2085965).
J.C.Y. acknowledges support from N.S.F Grant No.~AST-2108414.
This research has made use of the KMTNet system operated by the Korea Astronomy and Space 
Science Institute (KASI) and the data were obtained at three host sites of CTIO in Chile, 
SAAO in South Africa, and SSO in Australia.
\end{acknowledgements}


\begin{thebibliography}{}
\bibitem[Alard \& Lupton(1998)]{Alard1998} Alard, C., \& Lupton, R.\ H.\ 1998, \apj, 503, 325
\bibitem[Albrow(2017)]{Albrow2017} Albrow, M.\ 2017, MichaelDAlbrow/pyDIA: Initial Release on Github,Versionv1.0.0, Zenodo, doi:10.5281/zenodo.268049
\bibitem[Albrow et al.(2009)]{Albrow2009} Albrow, M., Horne, K., Bramich, D.~M., et al.\ 2009, \mnras, 397, 2099
\bibitem[Alcock et al.(1993)]{Alcock1993} Alcock, C. Akerlof, C. W., Allsman, R. A., et al.\ 1993, Nature, 365, 621
\bibitem[An(2005)]{An2005} An, J.~H.\ 2005, \mnras, 356, 1409
\bibitem[Beaulieu et al.(2016)]{Beaulieu2016} Beaulieu, J.-P., Bennett, D. P., Batista, V., et al.\ 2016, \apj, 824, 83
\bibitem[Bennett et al.(2010)]{Bennett2010} Bennett, D. P., Rhie, S. H., Nikolaev, S., et al.\ 2010, \apj, 713, 837
\bibitem[Bennett et al.(2016)]{Bennett2016} Bennett, D. P., Rhie, S. H., Udalski, A., et al.\ 2016, \aj, 152, 125
\bibitem[Bennett et al.(2018)]{Bennett2018} Bennett, D. P., Udalski, A., Han, C., et al.\ 2018, \aj, 155, 141
\bibitem[Bensby et al.(2013)]{Bensby2013} Bensby, T., Yee, J.~C., Feltzing, S., et al.\ 2013, \aap, 549, A147
\bibitem[Bessell \& Brett(1988)]{Bessell1988} Bessell, M.~S., \& Brett, J.~M.\ 1988, \pasp, 100, 1134
\bibitem[Bond et al.(2001)]{Bond2001} Bond, I. A., Abe, F., Dodd, R. J., et al. 2001, \mnras, 327, 868
\bibitem[Bozza(1999)]{Bozza1999} Bozza, V.\ 1999, \aap, 348, 311
\bibitem[Dan\v{e}k \& Heyrovsk\'y(2015)]{Danek2015} Dan\v{e}k, K., \& Heyrovsk\'y, D. 2015, \apj, 806, 99
\bibitem[Dan\v{e}k \& Heyrovsk\'y(2019)]{Danek2019} Dan\v{e}k, K., \& Heyrovsk\'y, D. 2019, \apj, 880, 72
\bibitem[Dominik(1998)]{Dominik1998} Dominik, M.\ 1998, \aap, 329, 361
\bibitem[Dominik(1999)]{Dominik1999} Dominik, M.\ 1999, \aap, 349, 108
\bibitem[Dong et al.(2009)]{Dong2009} Dong, S., Gould, A., Udalski, A., et al.\ 2009, \apj, 695, 970
\bibitem[Einstein(1936)]{Einstein1936} Einstein, A.\ 1936, Science, 84, 506
\bibitem[Gaudi et al.(2008)]{Gaudi2008} Gaudi, B. S., Bennett, D. P., Udalski, A., et al.\ 2008, Science, 319, 927
\bibitem[Gaudi \& Gould(1997)]{Gaudi1997} Gaudi, B.~S., \& Gould, A.\ 1997, \apj, 486, 85
\bibitem[Gould \& Loeb(1992)]{Gould+Loeb1992} Gould, A., \& Loeb, A.\ 1992, \apj, 396, 104 
\bibitem[Gaudi(1998)]{Gaudi1998} Gaudi, B. S., Naber, R. M., \& Sackett, P. D.\ 1998, \apj, 502, L33
\bibitem[Gould(1992)]{Gould1992} Gould, A.\ 1992, \apj, 392, 442
\bibitem[Gould(1994)]{Gould1994} Gould, A. 1994, \apj, 421, L75
\bibitem[Gould et al.(2010)]{Gould2010} Gould, A., Dong, S., Gaudi, B. S., et al.\ 2010, \apj, 720, 1073
\bibitem[Griest \& Safizadeh(1998)]{Griest1998} Griest, K., \& Safizadeh, N.\ 1998, \apj, 500, 37
\bibitem[Han(2005)]{Han2005} Han, C.\ 2005, \apj, 629, 1102
\bibitem[Han(2006)]{Han2006} Han, C.\ 2006, \apj, 638, 1080
\bibitem[Han et al.(2021a)]{Han2021a} Han, C., Albrow, M. D., Chung, S.-J., et al. 2021a, \aap, 652, A145
\bibitem[Han et al.(2019)]{Han2019} Han, C., Bennett, D. P., Udalski, A., et al.\ 2019, \aj, 158, 114
\bibitem[Han et al.(2001)]{Han2001} Han, C., Chang, H.-Y., An, J.~H., \& Chang, K.\ 2001, \mnras, 328, 986 
\bibitem[Han et al.(2005)]{Han2005} Han, C., Gaudi, B.~S., An, J.~H., \& Gould, A.\ 2005, \apj, 618, 962 
\bibitem[Han \& Gould(1995)]{Han1995}  Han, C., \& Gould, A.\ 1995, \apj, 447, 53
\bibitem[Han \& Gould(2003)]{Han2003}  Han, C., \& Gould, A.\ 2003, \apj, 592, 172
\bibitem[Han et al.(2022a)]{Han2022a} Han, C., Gould, A., Bond, I. et al.\ 2022a, \aap, in press 
\bibitem[Han et al.(2022b)]{Han2022b} Han, C., Gould, A., Kim, D., et al.\ 2022b, \aap, in press
\bibitem[Han et al.(2022c)]{Han2022c} Han, C., Udalski A., Lee, C.-U., et al.\ 2022c, \aap, 658, A93
\bibitem[Han et al.(2020a)]{Han2020-1953} Han, C., Kim, D., Jung, Y.~K., et al.\ 2020a, \aj, 160, 17 
\bibitem[Han et al.(2020b)]{Han2020-1700} Han, C., Lee, C.-U., Udalski, A., et al.\ 2020b, \aj, 160, 17 
\bibitem[Han et al.(2021b)]{Han2021b} Han, C., Lee, C.-U., Ryu, Y.-H. 2021b, \aap, 649, A91
\bibitem[Han et al.(2013)]{Han2013} Han, C., Udalski, A., Choi, J.-Y., et al.\ 2013, \apj, 762, L28
\bibitem[Han et al.(2017)]{Han2017} Han, C., Udalski, A., Gould, A., et al.\ 2017, \aj, 154, 223
\bibitem[Holtzman et al.(1998)]{Holtzman1998} Holtzman, J. A., Watson, A. M., Baum, W. A., et al.\ 1998, \aj, 115, 1946
\bibitem[Jung et al.(2021)]{Jung2021} Jung, Y.~K., Han, C., Udalski, A., et al.\ 2021, \aj, 161, 293
\bibitem[Jung et al.(2017)]{Jung2017} Jung, Y. K., Udalski, A., Bond, I. A., et al. 2017, \apj, 841, 75
\bibitem[Jung et al.(2018)]{Jung2018} Jung, Y. K., Udalski, A., Gould, A., et al.\ 2018, \aj, 155, 219
\bibitem[Kervella et al.(2004)]{Kervella2004} Kervella, P., Th\'evenin, F., Di Folco, E., \& S\'egransan, D.\ 2004, \aap, 426, 29
\bibitem[Kim et al.(2016)]{Kim2016} Kim, S.-L., Lee, C.-U., Park, B.-G., et al.\ 2016, J.~Kor. Astron. Soc., 49, 37
\bibitem[Lee et al.(2008)]{Lee2008} Lee, D.-W., Lee, C.-U., Park, B.-G., Chung, S.-J., Kim, Y.-S., Kim, H.-I., \& Han, C.  2008, \apj, ApJ, 672, 623 
\bibitem[Mao \& Paczy\'nski(1991)]{Mao1991} Mao, S., \& Pacz\'ynski, B.\ 1991, \apj, 374, L37 
\bibitem[Nataf et al.(2013)]{Nataf2013} Nataf, D.~M., Gould, A., Fouqu\'e, P., et al.\ 2013, \apj, 769, 88
\bibitem[Nemiroff \& Wickramasinghe(1994)]{Nemiroff1994} Nemiroff, R.~J. \& Wickramasinghe, W.~A.~D.~T.\ 1994, \apj, 424 , L21 
\bibitem[Robin et al.(2003)]{Robin2003} Robin, A. C., Reyl\'e, C., Derri\'ere, S., \& Picaud, S. 2003, \aap, 409, 523
\bibitem[Ryu et al.(2020)]{Ryu2020} Ryu, Y.-H., Udalski, A., Yee, J.~C., et al.\ 2020, \aj, 160, 183
\bibitem[Toonen et al. (2016)]{Toonen2016} Toonen, S., Hamers, A., \& Portegies Zwart, S. 2016, Computational Astrophysics and Cosmology, 3, 36
\bibitem[Udalski et al.(2005)]{Udalski2005} Udalski, A., Jaroszy\'nski, M., Paczy\'nski, B., et al.\ 2005, \apj, 628, L109 
\bibitem[Udalski et al.(1993)]{Udalski1993} Udalski, A., Szyma\'nski, M., Kalu\.{z}ny, J., et al. 1993, Acta Astron., 43, 289
\bibitem[Witt \& Mao(1994)]{Witt1994} Witt, H.~J., \& Mao, S.\ 1994, \apj, 430, 505
\bibitem[Yee et al.(2012)]{Yee2012} Yee, J.~C., Shvartzvald, Y., Gal-Yam, A., et al.\ 2012, \apj, 755, 102
\bibitem[Yoo et al.(2004)]{Yoo2004} Yoo, J., DePoy, D.~L., Gal-Yam, A., et al.\ 2004, \apj, 603, 139
\bibitem[Zang et al.(2021a)]{Zang2021a} Zang, W., Han, C., Kondo, I., et al.\ 2021, Res. in Astro. and Astrophys., 21, 239
\vspace*{\fill}
\end{thebibliography}
\end{document}